\documentclass[11pt]{article}
\usepackage{helvet}
\usepackage{geometry}                % See geometry.pdf to learn the layout options. There are lots.
\usepackage[parfill]{parskip}    % Activate to begin paragraphs with an empty line rather than an indent
\usepackage{graphicx}
\usepackage{amssymb}
\usepackage{amsmath}
\usepackage{epstopdf}

\usepackage{cite}
\title{\bf Vapor-liquid-solid growth of serrated GaN nanowires: Shape selection driven by kinetic frustration}
\author{\bf Zheng Ma$^1$, Dillon McDowell$^1$, Eugen Panaitescu$^1$, Albert V. Davidov$^{2}$, \\
\bf Moneesh Upmanyu$^{3,4}$, and Latika Menon$^1$\\
\small $^1$Department of Physics, Northeastern University, Boston, MA 02115\\
\small $^{2}$Material Measurement Laboratory, National Institute of Standards and Technology, \\
\small Gaithersburg, Maryland 20899, USA\\
\small $^{3}$Group for Simulation and Theory of Atomic-Scale Material Phenomena ({\it st}AMP), \\
\small Department of Mechanical and Industrial Engineering, \\
\small Northeastern University, Boston, MA 02115\\
\small $^{4}$Department of Bioengineering, Northeastern University, Boston, MA 02115
}

\date{}                                           % Activate to display a given date or no date

\begin{document}
\maketitle

\noindent
{\bf Abstract: Compound semiconducting nanowires are promising building blocks for several  nanoelectronic devices yet the inability to reliably control their growth morphology is a major challenge. Here, we report the Au-catalyzed vapor-liquid-solid (VLS) growth of GaN nanowires with controlled growth direction, surface polarity and surface roughness. We develop a theoretical model that relates the growth form  to the kinetic frustration induced by variations in the V(N)/III(Ga) ratio across the growing nanowire front. The model predictions are validated by the trends in the as-grown morphologies induced by systematic variations in the catalyst particle size and processing conditions. The principles of shape selection highlighted by our study pave the way for morphological control of technologically relevant compound semiconductor nanowires.}

\pagebreak

The vapor-liquid-solid (VLS) route for high-yield nanowire synthesis  allows control over the growth form and composition via a capping catalyst particle that serves both as a catalyst and conduit for material transfer onto the growing wire\cite{nw:WagnerEllis:1964, nw:Lieber:1998, nw:DuanLieber:2000, nw:CuiLieber:2001, Dubrovskii2006}. The size of the particle is the primary length-scale that controls the nanowire diameter while the time-scale is usually set by a combination of slow surface catalysis and nucleation at the catalyst-nanowire interface\cite{nw:KodambakaRoss:2006, nw:KimRoss:2008, nw:Ross:2010}. 
%Then, aspects related to the growth morphology, in particular the growth direction and the shape, are largely determined by the interplay between the supersaturation within the particle and energetics of the particle-nanowire interphase as well as the exposed nanowire surfaces. 
In the case of compound semiconductors, the growth morphology of thin films is strongly influenced by the relative rates of incorporation of the constituent species\cite{tsf:SchwaigerMetzner:2011}, and such effects must also modify growth of nanowires. Indeed, recent studies on semiconducting III-V and II-VI nanowires have shown that the incorporation ratio has direct effect on their growth rate and crystal quality\cite{nw:ChenChen:2001, nw:DayehYuWang:2007, nw:LiTang:2009}, yet the fundamental mechanisms that affect the growth energetics and kinetics remain unknown. In this article, we systematically explore this interplay during Au-catalyzed growth of GaN nanowires, a wide band-gap material of direct relevance in optoelectronics and power nanoelectronics\cite{tsf:NakamuraSenoh:1994, nw:ChoiJohnson:2003, book:Morkoc:2007}. The combination of experiments and theoretical frameworks identify a novel mechanism based on interplay between surface energetics and catalytic kinetics for shape selection in GaN with broad implications for controlled growth of vertical arrays of compound nanowires.

\begin{figure}
\centering
\includegraphics[width=\columnwidth]{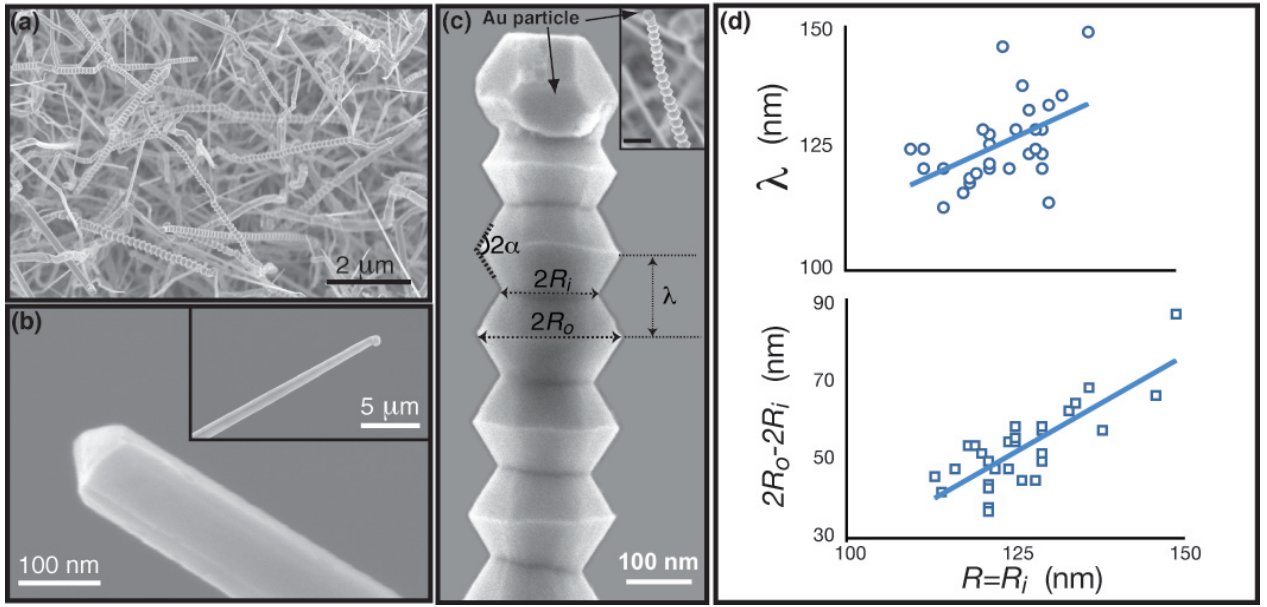}
\caption{\label{fig:figure1} {\bf \small Morphology of as-synthesized GaN nanowires}: (a) SEM characterization of Au-catalyzed GaN nanowires grown via chemical vapor deposition at temperature $T=960\,^\circ$C and pressure $P=100$\,Torr. The Ga partial pressure corresponds to the initial mass $m=1.02$\,g  of the solid precursor Ga$_2$O$_3$. (b) High-resolution image of a straight nanowire with the exposed growth front that reveals a triangular cross-section. (inset) Low-resolution image showing the smooth lateral facets and the capping catalyst particle. (c)  Low- (inset, scale bar $200$\,nm) and high-resolution image of a serrated GaN nanowire capped by an Au-catalyst particle. The parameters associated with the nanowire morphology are as indicated: $2R_i=125$\,nm $2R_o=200$\,nm, $\lambda=120$\,nm, and $\alpha\approx60^\circ$.  (d) Size dependence of the wavelength $\lambda$ (top) and amplitude $2R_o-2R_i$ (bottom) associated with the serrations.
}
\end{figure}

The nanowires are grown by ammoniating solid Ga$_2$O$_3$ over an Au-decorated $\langle111\rangle$ Si substrate at $T=960\,^\circ$C and pressure $P=100$\,Torr (see Methods). The use of hydrogen as a carrier gas builds Ga$_2$O$_3$ vapor pressure under these conditions that is readily abstracted by the flowing NH$_3$ gas to elemental Ga and N\cite{tsf:Taylor:1999, nw:NamFischer:2004, nw:WuMenon:2009, nw:TsivionJoselevich:2011}. The growth is evident in the low-resolution SEM image (Fig.~\ref{fig:figure1}a). Almost all nanowires are capped by a Au particle characteristic of the classical vapor-liquid-solid (VLS) growth mechanism. Uncatalyzed vapor-solid (VS) lateral growth of the nanowires is much slower and effectively suppressed under these conditions\cite{nw:ChengMao:1999, nw:LiWang:2000, nw:PengLee:2000, nw:NamFischer:2004, nw:BertnessDavydov:2008}, a fact further validated by negligible diameter coarsening upon varying the exposure time (not shown). X-ray diffraction of the as-grown nanowires confirms that they are strain-free hexagonal wurtzite crystals (Supplementary Figure~S1). 
%This is consistent with the relatively high growth temperature employed in this study. 

%\subsection*{Straight nanowires}

The as-grown morphology consist of straight nanowires with smooth sidewalls as well as several instances of rough yet highly periodic surface morphologies. Higher resolution SEM images of the smooth nanowires are shown in Fig.~\ref{fig:figure1}b. The nanowire radius is approximately that of the catalyst particle during axial growth (inset). In some cases, the droplet etches away and exposes a triangular nanowire cross-section, evident in the figure. Lattice fringes with inter-planar distance $d\approx2.78$\,{\rm \AA} are visible in HRTEM image normal to the growth axis (Supplementary Figure~S2). Secondary fringes aligned $30^\circ$ to the wire axis are also evident as indicated. We rarely see any interruptions in the fringes indicating that the nanowires are largely defect-free. Taken {\it in toto}, the nanowire is composed of non-polar $m$-planes stacked along the $\langle10\bar{1}0\rangle$ direction. For a nanowire so grown, one of the three lateral facets is the polar $\{0001\}$ plane and the other two belong to the $\langle11\bar{2}2\rangle$ family of directions. The non-polar growth is classical in that it has been reported under a variety of conditions and several synthesis techniques\cite{nw:BaeSangsig:2003}, including the VLS route\cite{nw:KuykendallPauzauskie:2003, nw:KuykendallPauzauskie:2004}.

\begin{figure}
\centering
\includegraphics[width=\columnwidth]{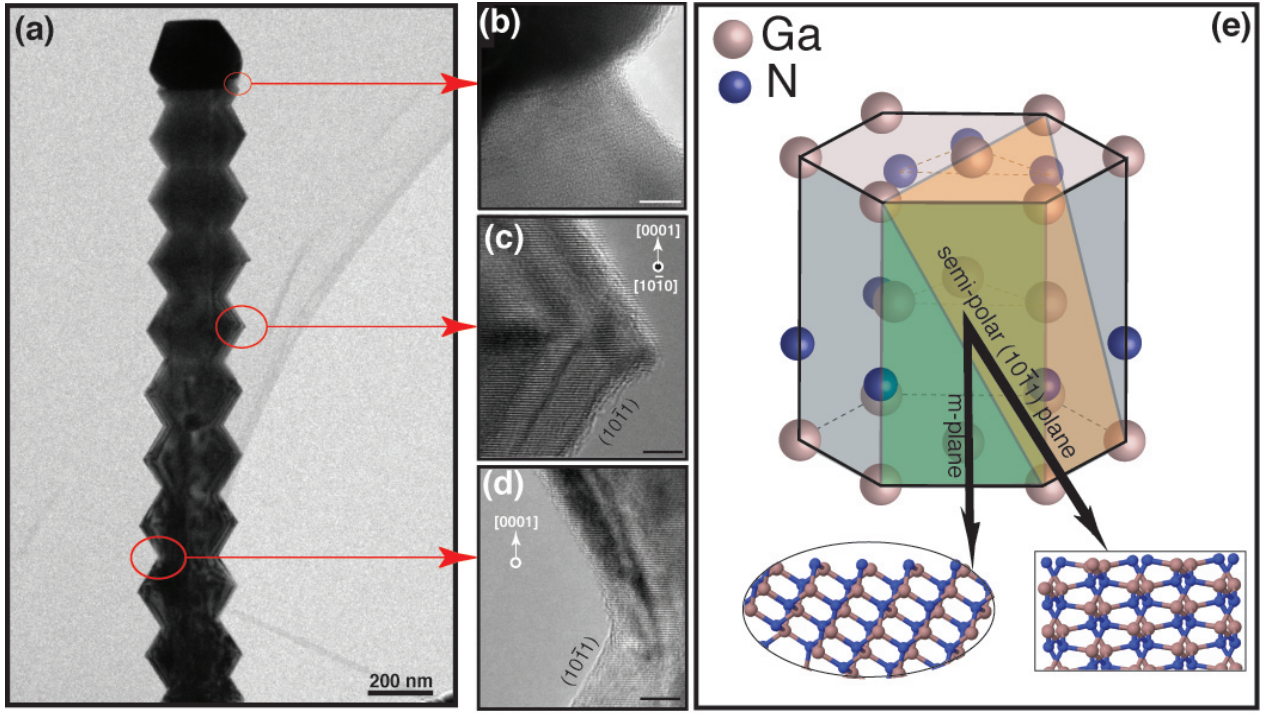}
\caption{\label{fig:figure2} {\bf \small Characterization of serrated nanowires}: (a) Low-resolution TEM image of a serrated nanowire. (b) The magnified view of the particle-nanowire interface near the contact line. The image is tilted slightly with respect to (a) for clarity. (c-d) Magnified views of the intersection of the two facets as indicated by the arrows. The scale bar is 5 nm. The fringes in (c) and (d) correspond to planes normal to the growth direction. The average inter-planar spacing of $d=2.63$\,{\rm \AA} is consistent with growth along the polar $[0001]$ direction. (e) The hexagonal wurtzite GaN unit cell. The non-polar $m$-plane (shaded green) and the semi-polar $\{10\bar{1}1\}$ plane (shaded yellow) are shown as reference. The insets show the unreconstructed semi-polar surface structure observed along two orthogonal directions indicated by arrows.}
\end{figure}

%\subsection*{Serrated nanowires}
Of interest is the serrated morphology and the conditions that stabilize its growth. Majority of these serrated nanowires have larger diameter. The SEM images in Fig.~\ref{fig:figure1}c show the detailed view of the nanowire morphology. The growth of these nanowires is also via the VLS mechanism as the nanowire is always capped with the Au particle. 
%The Au particle can sometimes migrate to the sidewalls, as is the case here. 
Figure~\ref{fig:figure2}a shows another example with a stable Au particle. Although the catalyst particle is faceted at room temperature, the high growth temperature Au-Ga solution forms a melt for Ga concentrations above a critical value. This follows from the phase diagram which predicts a stable liquid Au-Ga solution for Ga concentrations above $5$\,at\%\cite{pd:ElliottShunk:1981, pd:OkamotoMassalski:1984}. The critical composition is size dependent yet the growth form is unaffected as the corrections simply scale the critical Ga composition.  

The size of the catalyst particle is approximately the average nanowire size defined as $R_i$. The low-resolution image in the inset shows that the nanowire maintains its hexagonal cross-section.
%suggesting that the growth occurs along the $c$-axis. 
The lateral facets, visible in the the magnified image are consistent with the six-fold symmetry and give an appearance of truncated hexagonal bipyramids stacked along the growth direction. The  serrations are remarkably periodic; for a given nanowire radius $R_i$ the wavelength $\lambda$ and the amplitude $2A=(R_o-R_i)$ are both preserved along the length of the nanowire. The facet reorientation angle $2\alpha$ is also preserved but it is independent of the nanowire size, i.e. $2\alpha\approx120^\circ$ averaged several nanowire segments.

HRTEM characterization of the serrations shown in Fig.~\ref{fig:figure2} do not indicate the presence of any obvious defects/grain boundaries. Rather, the lattice fringes are unbroken and continue across the serrations indicating that the nanowire is a high quality single crystal. This crystalline nature is also preserved at points where the sidewall facets reorient (Fig.~\ref{fig:figure2}c and Fig.~\ref{fig:figure2}d). Viewed approximately normal to the growth axis we see lattice fringes with clear variations in grayscale contrast. The fringe spacing $d\approx2.63$\,{\rm\AA} is consistent with growth of polar bilayers along the $\langle0001\rangle$ direction ($c$-axis). Unlike classical polar growth, though, the enveloping sidewall facets are inclined at an angle $\alpha\approx60^\circ$ for each facet which we then identify to be the $\{10\bar{1}1\}$ or $\{11\bar{2}2\}$ family of planes. Selective area electron backscatter diffraction (EBSD) analysis of serrated segments confirms the growth direction and identifies the orientation of the semi-polar planes that form the sidewall facets of the truncated hexagonal bipyramid segments to be the $\{10\bar{1}1\}$ family of planes; the analysis is detailed in Supplementary Figure 3. 
%As confirmation, we also see lattice fringes parallel to the surface spaced at a distance $d\approx0.271$\,nm (Supplementary Information), in excellent agreement with the bilayer spacing for these semi-polar planes.  
The wurtzite unit cell drawn schematically in Fig.~\ref{fig:figure2}e shows the details of the crystallographic relationship for one of the six semi-polar planes. The atomic configurations of the (unreconstructed) N-polar surface structure viewed along two orthogonal directions are also shown in the insets. 
\begin{figure}
\centering
\includegraphics[width=0.8\columnwidth]{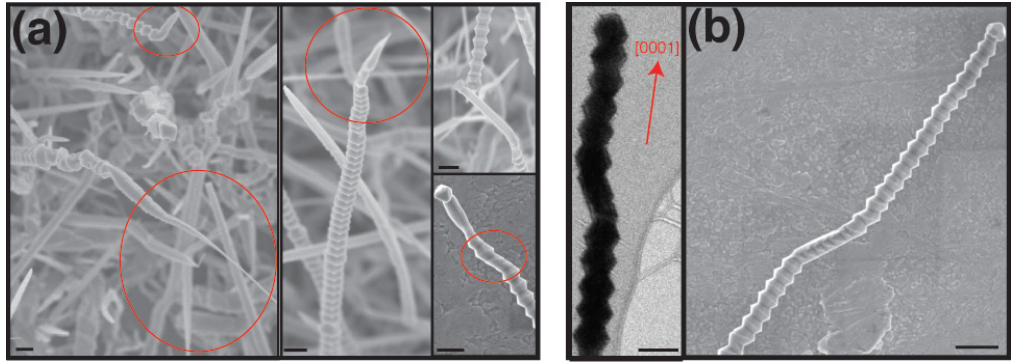}
\caption{\label{fig:figure3} {\bf \small Variations and transitions in GaN nanowire growth morphology}: (a) SEM images showing serrated-to-smooth shape transitions. The scale bar is 200\,nm. (left and middle) Circled regions indicating distinct reduction in the nanowire diameter accompanying the transition. (top and bottom, right) Early stages of the transition to a non-polar growth direction with smooth sidewall morphology. The encircled region (bottom) shows the discrete kinks consisting of truncated bipyramids. (b) Local and reversible morphology changes seen in low-resolution TEM image (left) and high resolution SEM images (top right). 
%(bottom right) Detailed view of the transition zone showing the kinks. 
%(c) Schematic showing the main elements of the theoretical model that relates the size and the III/V ratio to the growth morphology of the semi-polar $\{11\bar{2}1\}$ nanowires. See text and SI for details. The nitrogen incorporation is assumed to occur over an exposed annular ring (thickness $\delta R$) enveloping the contact line (shaded blue), which also serves as the preferred site for GaN nucleation as illustrated. (inset) Schematic illustration of the widening and narrowing morphology of the particle-nanowire system. 
}
\end{figure}

The serrated morphology depends on the nanowire size and the variation is plotted in Fig.~\ref{fig:figure2}d. Although there is scatter, both $\lambda$ and $2A$ increase with the nanowire size $R_i$. A linear fit to our data yields $\lambda\approx0.85R_i$. The amplitude of the serrations varies as $2A\approx R_i$ and yields the wavelength/amplitude ratio, $\lambda/A\approx1.7$. The ratio is remarkably small indicating a fully faceted yet extremely rough morphology which can be tuned simply by changing the nanowire size.

A careful examination of the as-grown samples also reveals transitions between the two growth morphologies. Several instances of serrated-to-smooth transitions  are highlighted in Fig.~\ref{fig:figure3}a (encircled, left and middle panels). We have also carried out EBSD analysis transition zones and the results are shown in Supplementary Figure 3. The change in growth morphology is preceded by a gradually tapering cross-section, i.e. the dynamic transitions are strongly correlated with decrease in the nanowire size. The tapering is most likely due to a decrease in the catalyst particle size during growth as the Au atoms diffuse down the nanowire sidewalls.\cite{nw:HannonRossTromp:2006} In each case, the effective nanowire growth direction changes (top right). Closer examination shows that the transition point  consists of a series of visibly discrete kinks consisting of increasingly truncated bipyramids such that the effective growth direction is along intermediate, low-energy orientations (right, bottom). Although the crystallographic orientation of each of the kinks is still polar, the effective orientation of these transition zones is semi-polar. In essence, the kinking reflects local variations in processing conditions and/or geometry that modify the growth direction as well as the nanowire morphology. 
 
We see similar kinks within confined regions in an otherwise serrated nanowire (Fig.~\ref{fig:figure3}b). Evidently, the change in the growth direction is not dynamically stable as the nanowire reverts back to its original growth direction. There is negligible change in the nanowire size (left and top right panels) and detailed characterization of the transition zone shows modified serrations due to the intermediate semi-polar growth orientation/s (bottom right panel). We note that majority of the transition are observed at relatively large Ga partial pressures (higher mass of solid Ga$_2$O$_3$), and we do not observe smooth-to-serrated transitions.
 
%\section*{Growth model} 
The interplay between the growth direction and the V:III ratio is consistent with several past studies on GaN thin films and nanowires\cite{tsf:KoleskeGorman:1998, tsf:HierroSpeck:2002, nw:NamFischer:2004}. The non-polar growth direction is favored for ratios close to unity. This indicates that the combination of growth parameters that result in growth of the $\langle10\bar{1}0\rangle$ non-polar nanowires is such that that the V:III ratio at the growth front is close to unity. Reducing the amount of Ga in the gas phase increases the V:III ratio at the growth front and the growth direction switches to the polar direction. 

The serrated morphology of the polar nanowires is unlike past reports of the rough morphologies observed during VS growth. The surface roughness there is controlled by the Ga diffusion length that is directly influenced by the V:III ratio; low ratios result in smooth morphologies while large ratios lead to surface roughening as the Ga diffusion length at the crystal-vapor interface is dramatically reduced by N-rich conditions\cite{nw:NamFischer:2004}. This is in direct contrast to the our observations as the serrations occur under N-deficient conditions. Evidently, the droplet and the solid-liquid interface plays a central role in the combined interplay between the size, interface energetics and kinetics during the morphological evolution of the nanowire. 

\begin{figure}
\centering
\includegraphics[width=0.75\columnwidth]{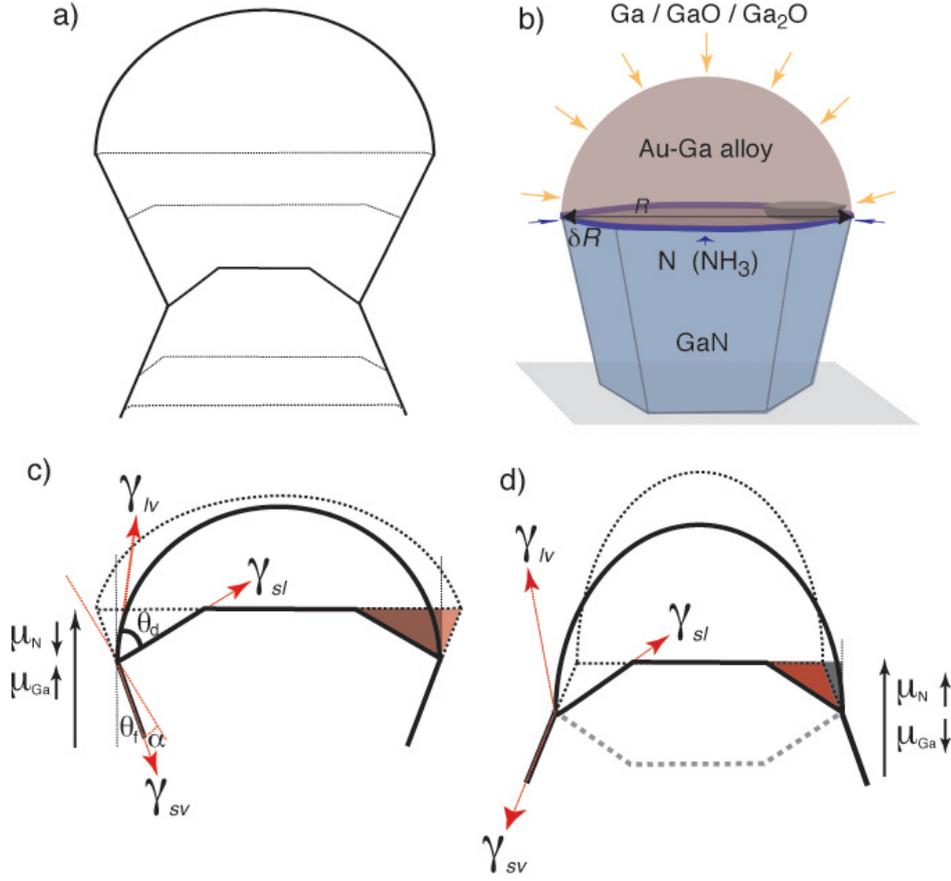}
\caption{{\label{fig:figure4} \small \bf Growth model}: (a) Schematic showing the evolution of the droplet-nanowire system as the size oscillates. The interplay between size dependent droplet supersaturation and the fully faceted morphology of the droplet-nanowire interface is shown. (a) Schematic showing the main elements of the theoretical model that relates the size and the V:III ratio to the growth morphology of the serrated nanowire. The nitrogen incorporation is assumed to occur over an exposed annular ring (thickness $\delta R$) enveloping the contact line (shaded blue), which also serves as the preferred site for GaN nucleation as illustrated. (c-d) The energetics at the contact line during (c) the widening (d) and narrowing phases of the growth. The colored regions near the contact line represent the volume swept out by the truncating facets during each oscillatory growth cycle. The dark solid and dotted lines represent the initial and final states of the droplet and the sidewall facets, respectively. The lightly shaded dotted line in (d) is an alternate concave morphology of the solid-liquid interface that is also possible as the sidewalls narrow inwards. See text for details.
}
\end{figure}

Development of a detailed theoretical model is handicapped by the lack of quantitative understanding of these aspects in GaN in particular and multicomponent systems in general, and we present a simplified analysis of the competing effects. Current understanding of the interplay between size and energetics during general VLS growth revolves around the interfacial balance at the vapor-particle-nanowire trijunction\cite{nw:GivargizovChernov:1973, nw:Givargizov:1975, nw:DubrovskiNickolai:2004, nw:KodambakaRoss:2006, nw:KodambakaRoss:2007, nw:RoperVoorhees:2007, nw:SchwalbachVoorhees:2008, nw:SchwarzTersoff:2011, nw:SchwarzTersoff:2011b}. The serrations are  reminiscent of similar growth form observed during VLS growth of Si(111) nanowires marked by absence of low-energy facets normal to the growth direction. Then, surface energy considerations alone force the nanowire size to shrink or widen to accommodate the low-energy inclined facets as illustrated schematically in Fig.~\ref{fig:figure4}a. The size cannot change monotonically since it is constrained by the wetting angle between the droplet and the nanowire along the contact line. As the nanowire widens or narrows, the droplet is stretched  and compressed respectively. Beyond a critical size, the pressure exerted by droplet forces the sidewalls to reorient at regular intervals. Ross {\it et al.} have recently analyzed this interplay between geometry and energetics for 2D near-equilibrium nanowire growth based on a balance between i) changes in the surface and (nanowire-particle) interphase area, and ii) the work done against the surface tension $\gamma_d$ of an almost hemispherical particle\cite{nw:RossTersoffReuter:2005}. The angle $\theta_f$ between the facets and the growth axis is fixed as the shrinking and growing facets belong to the same family of planes. Then, the wavelength of the serrations varies as\cite{nw:RossTersoffReuter:2005}
\begin{equation}
\lambda \sim  \frac{\Gamma_c}{\gamma_d\,\theta_f\sin\alpha}\;R_i\;,
\end{equation}
where $\Gamma_c$ is a critical barrier associated with facet reorientation at the apex and troughs in the serrations. 
The size dependence of the serration wavelength and amplitude plotted  is roughly consistent with this theoretical model. The small $\lambda/A$ ratio indicates that the serrations are large in extent and sensitive to the size suggesting a large barrier $\Gamma_c$. 

There is considerable scatter in the size dependence, though, and that is likely due to several oversimplying assumptions. One, the model assumes that the nanowire-particle interface is flat. Recent studies have shown that that is not the case, rather the  interface is enveloped by truncating facets as it meets the contact line\cite{nw:OhChisholmRuhle:2010, nw:WenTersoffRoss:2011, nw:RyuCai:2011, nw:WangUpmanyu:2013}. The fully faceted morphology is shown schematically in Fig.~\ref{fig:figure4}a. Note that the equilibrium Wulff plot associated with the inclination dependent solid-liquid interface energy determines the relative areas of the two classes of facets. Following Ref.~[35], the change in energy of the nanowire per unit length of growth becomes
\begin{align}
\label{eq:energy}
\frac{dE}{dz} = R\Delta\mu + \frac{\gamma_{sv}}{\cos\theta_f} + (\gamma_d\cos\theta_d + \gamma_{sl}^t)\sin\alpha + \gamma_{sl}^m \sin(\alpha+\theta_f)
\end{align}
where $\theta_f+\alpha$ is the angle that the truncating facet makes with the main facet. The first term is the energy gain due to the driving force of the excess chemical potential, the second term is the work done against the liquid surface tension, and the last two terms are the costs associated with increasing the areas of the truncating and main facets with energies $\gamma_{sl}^t$ and $\gamma_{sl}^m$, respectively. For small $\alpha$ such that the truncating facet is approximately normal to the sidewalls, the energy minimizing morphology is one with low $\gamma_{sv}$ and $\gamma_{sl}^m$.  The droplet surface tension plays a minor role and this highlights another factor that stabilizes the truncating facet and also determines its orientation. It has not escaped our attention that the small $\alpha$ assumptions require that solid-liquid interface must adopt a concave morphology, i.e. the truncating facets switch their orientation about the horizontal. This can certainly take place during the reorientation of the sidewall facets. The concavity has implications for the supersaturation of the droplet and this is addressed below.

The second assumption relates to the the chemical potential difference $\Delta\mu$ between the solid nanowire and supersaturated droplet that drives the nanowire growth. The is a key issue as N has negligible solubility within the Au particle\cite{pd:OkamotoMassalski:1984} and its catalysis is limited to the contact line, approximated as an annular ring of width $\delta R$. Ga incorporation, on the other hand, occurs through the particle as Au not only catalyzes the decomposition of the Ga precursor but also forms a stable Au-Ga solution\cite{pd:ElliottShunk:1981}. The scenario is shown schematically in Fig.~\ref{fig:figure4}b.
%At low Ga concentration upto the solvus $X^{Ga}_{eq}\approx1\%$, the particle is a solid solution while above the liquidus composition $X^{Ga}_{eq}\approx8\%$ it forms a liquid solution.\symbolfootnote[2]{Note that we have ignored size effects which can modify the equilibrium concentrations} 
Then, the size of the Au-particle impacts the V:III ratio at the growth front in that the atomic incorporation rates $\mathcal{I}^{Ga}$ and $\mathcal{I}^N$ scale differently with the catalyst particle size, $R$. Ignoring differences in the atomic volume of Ga and N across the phases, the steady-state supersaturation at the growth front evolves as (see Supplementary Notes, Eqs.~S1-S8)
\begin{align}
\label{eq:eq2}
\frac{d X^{Ga}_l}{dz} = \frac{d(X^{Ga}_l-X^{Ga}_{l(eq)})}{dz} & \sim \frac{1}{vR}\left(k^{Ga}_{vl} -v\right)\\
\label{eq:eq3}
\frac{d X^{N}_{c}}{dz} \approx  \frac{d X^{N}_c}{dz} &\sim  \frac{k^N_{c}}{v}  - \frac{R}{\delta R},
\end{align}
where $v$ is the overall nanowire growth velocity and $k^{Ga}_{vl}$ and $k^N_c$ are the catalysis rate constants for Ga and N at the particle (liquid-vapor) surface and contact line, respectively. The first term in each relation is the build up due to catalysis and the second term is the decrease due to nanowire growth.  In the limit that the growth velocity is size-independent\cite{nw:KodambakaRoss:2006, Dubrovskii2007, nw:SchmidtGosele:2007}, the excess N along the contact line decreases linearly with nanowire size. The decrease is quite dramatic since we expect the width of the contact line to be order of the interatomic width\footnote{The droplet can control this width by receding away from the nanowire edge. We ignore this effect, although such a regulatory mechanism has been reported for epitaxially grown GaN nanowires\cite{nw:KuoChen:2006}.}. 

The scaling relations (\ref{eq:eq2} and \ref{eq:eq3}) capture the main effect of size on the asymmetry in the build up of excess Ga and N and therefore the V:III ratio at the growth front. This has a direct effect on the driving force as both the Ga and N chemical potentials are related to their supersaturation. Consider the extreme case where the N-supply sets the overall nanowire growth velocity. Then, the excess N-coverage along the contact line is negligible such that $d X^{N}_{c}/dz\approx0$ and Eq.~\ref{eq:eq3} reduces to 
\begin{align}
v\sim\frac{\delta R}{R}  k^N_c.
\end{align}
%where $\ddot{G}(0)$ is the second derivative of the free energy associated with the N-coverage at equilibrium ($X^N_c=0$). 
Then, the periodic modulations in the nanowire size during serrated growth are coupled to oscillations in Ga supersaturation in the droplet and therefore the overall V:III ratio at the growth front (Supplementary Eq. 16). Specifically,
\begin{align}
\frac{d(\Delta \mu^{Ga})}{dz} &\sim  \frac{\ddot{G}(X^{Ga}_{l(eq)})}{vR} \left(k_{vl}^{Ga} - \frac{\delta R}{R}k_{c}^N  \right) + \frac{d\sum_i\kappa^\gamma_i}{dz},\nonumber\\
\frac{d\Delta \mu^{N}}{dz} &\sim \gamma_d \cos\theta_d \frac{d\theta_d}{dz}.
\end{align}
The second term in $\Delta\mu^{Ga}$ is the contribution from the weighted mean curvatures of the two classes of facets that make up the nanowire-particle interface\cite{cg:Taylor:1992b, cg:CarterCahnTaylor:1995}. It is directly related to the relative lengths (areas) of the facets. In essence, its effect is such that chemical potential in the droplet varies inversely with the extent of truncation\cite{nw:WenTersoffRoss:2011}. Observe that $\mu^N_c$ is regulated by the droplet contact angle. The combined effect results in a non-trivial interplay between the net driving force $\Delta \mu = X^{Ga}_{GaN} \Delta\mu^{Ga}_l + X^N_{GaN} \Delta \mu^N_{c}$ and the energy minimizing sidewall facet given by minimization of Eq.~\ref{eq:energy}.

\begin{figure}
\centering
\includegraphics[width=0.8\columnwidth]{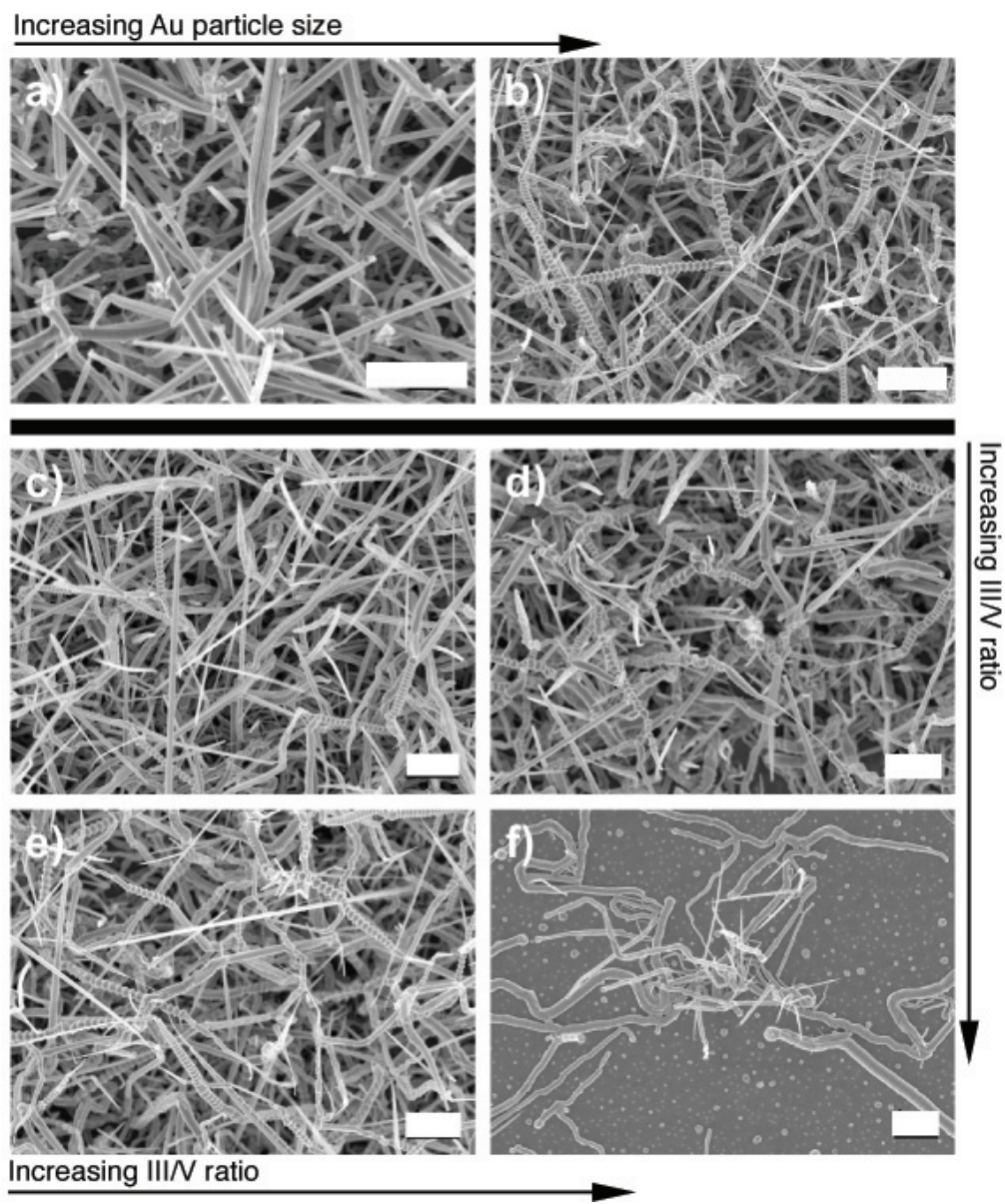}
\caption{\label{fig:figure5} \small {\bf Interplay between particle size and III/V ratio and growth morphology}: (a)-(b) Effect of Au catalyst size at fixed Ga$_2$O$_3$ partial pressure, $m=0.1.02$ g. (a) The serrated morphology is noticeably absent for small sizes ($R\approx 80$\,nm) and dominates much of the as-synthesized network at larger sizes  ($R>100$\,nm). (c-f) Effect of increasing partial pressure of Ga$_2$O$_3$; the input mass for the four cases are $m=0.61$\,g, $m=0.75$\,g, $m=1.03$\,g and $m=1.91$\,g, respectively.
}
\end{figure}
The analysis is consistent with recent experiments that have shown direct correlation between the V:III ratio and changes in nanowire size\cite{nw:LimGradecak:2013}. The size change is believed to be entirely due to change in droplet size, i.e. there the kinetics results in an interplay between the V:III ratio and the droplet size Unlike these studies, though, the highly periodic modulations we observe here indicate a self-regulating mechanism which follows directly from size controlled chemical potential variations within the liquid droplet. More concretely, the growth of large diameter nanowires is nitrogen deficient which initially leads to diameter reduction along the energy minimizing facet orientation of the sidewalls, i.e. the sidewall shape selection is no longer governed completely by interface energetics. It is now modified by kinetics since it must also increase the V:III ratio while maintaining growth along the $c$-axis. Note that if ignore the much smaller changes in the droplet size, the droplet angle $\theta_d$ increases as the droplet is squeezed in and this enhances the $\mu^N_c$. At a critical point, the V:III ratio increases to close to unity. The nanowire can change its growth direction to non-polar growth, and we see successful as well as failed attempts at these transitions in Fig.~\ref{fig:figure3}. Clearly, there is an energy barrier that needs to be overcome to effect the growth direction change. The barrier for changing the growth direction is expectedly smaller than that for reorientation of the sidewall facets, and the nanowire starts to widen by reorienting the sidewall facets which also alleviates the compression in the droplet.  At this point, the Ga supersaturation in the droplet is at a minimum as the difference in the effect the effective catalysis rates $k_{vl}^{Ga} - (\delta R/R)\, k_{c}^N $ decreases with size. Then, the nanowire-particle interface is significantly truncated as shown in Fig.~\ref{fig:figure4}a. The widening of the sidewalls again minimizes the interface energies and occurs along the direction that rectifies the Ga deficiency. The droplet becomes increasingly stretched and at a critical point the capillary forces favor narrowing of the sidewall facets. The nanowire again reorients. The droplet is significantly Ga-rich and that minimizes the lengths (areas) of the truncating facets. As mentioned earlier, the reorientation into the narrowing sidewalls can lead to a faceted yet concave morphology, as shown in Fig.~\ref{fig:figure4}d (lightly shaded). This not at odds with the Gibbs-Thomson effect associated with the solid-liquid interface as the droplet is significantly N-deficient and therefore {\it undercooled} with respect to the solid nanowire\cite{cg:CarterCahnTaylor:1995}. Although we do not have direct evidence during growth, \ref{fig:figure2}b does suggest a concave interfacial morphology near the contact line.

The kinetic anisotropy of the growth rates can also dictate the shape selection of the sidewall facets. This follows from the fact that non-polar planes normal to the to $c$-axis {\emph do} exist, and we generally expect these to be lower in energy compared to the semi-polar surfaces. This is by no means conclusive as it is well-known that relative energies are sensitive to the reactor conditions including the V:III ratio and the partial pressures in the gas phase. Nevertheless, selected area growth (SAG) experiments on $c+$ axis GaN crystals show that under N-deficient conditions, the growth is dominated by the very same semi-polar sidewalls\cite{tsf:HiramatsuMaeda:1999}. Current understanding of the shape selection during vapor-solid (VS) crystal growth is based on the {\it kinetic} Wulff-plots that can completely dominate the growth form for large kinetic undercoolings that scale as $\Delta \mu=v/M$, with $M$  the mobility of the growth front\cite{nw:DuSrolovitzMitchell:2005}. The serrated growth occurs through a Ga-rich droplet under N-deficient conditions, and therefore we expect  $c+$ (Ga) growth for which the $\langle10\bar{1}1\rangle$ planes have been observed to the slowest growing and therefore the dominate the sidewalls during growth of the convex and faceted solid-liquid interface. Interestingly, SAG experiments on annular rings reveal that the concave growth is dominated by $\langle11\bar{2}2\rangle$ semi polar facets since they are the fastest growing facets\cite{tsf:HiramatsuMaeda:1999}. Although there is overwhelming evidence that the sidewall facets in the serrated nanowires here are enveloped by the $\langle10\bar{1}1\rangle$ planes, we cannot completely rule out the presence of $\langle11\bar{2}2\rangle$ planes as they have approximately the similar inclination to the c-axis ($28^\circ$ versus $32^\circ$). Then, the inclination dependent kinetic anisotropy in the growth rates solid-liquid interface would become important. The sidewall orientation selection occurs during crystallization from the droplet, and more specifically during each oscillatory cycle of growth consisting of increase and decrease in the extent of the truncating facets which eventually leads to the nucleation of a new layer on the main facet\cite{nw:OhChisholmRuhle:2010, nw:WenTersoffRoss:2011}. The truncating facets maintain their orientation as they change their size and the contact line moves up and down the sidewalls during growth cycle. This is shown schematically in Fig.~\ref{fig:figure4}c and Fig.~\ref{fig:figure4}d. As reference, the path traced by the contact line for straight and narrowing/widening sidewalls are also shown. Then, for convex growth, the kinetic anisotropy will pick the $\langle10\bar{1}1\rangle$ provided it is the slowest moving solid-liquid interface inclination for the range of all possible orientations that preserve $c+$ polar growth.  

Although it is clear that kinetic frustration modifies the principles of shape selection in the serrated nanowires, a quantitative understanding of the anisotropic energetics and kinetics of solid-liquid and solid-vapor interfaces is essential to develop a more concrete understanding of GaN nanowire growth. To test several elements of our model, we have repeated our growth experiments with i) varying average Au seed particle sizes, and ii) varying V:III ratio. In each case, we report the yield of serrated nanowires over the entire growth region. Figure~\ref{fig:figure5} shows SEM images of the nanowires under varying growth conditions. A two-fold increase in the Au particle size results in an almost three-fold increase in the yield of serrated nanowires, showing a clear correlation between the growing particle size and the nanowire morphology. Similarly, decreasing the V:III ratio by increasing the mass of the precursor oxide results in significant increase in the yield of the serrated nanowires indicating the N-deficiency promotes the formation of the serrations. At very low ratios (\ref{fig:figure5}e), the growth is sporadic and marks an upper limit for VLS growth. More importantly, both trends are consistent with the main elements of the growth model, and we expect the interplay between size, component ratio and growth morphology to be a general aspect of VLS growth of multicomponent nanowires.

\section*{Discussion and Conclusions}
Our study highlights the potential of VLS route for concurrent control over the growth direction and morphology of multicomponent nanowires. The interplay between catalyst particle size and the V:III ratio can be further tuned by the controlling the growth temperature, and although this remains to be explored systematically, we expect a considerably rich set of growth morphologies. Control over the surface roughness has important ramifications for their properties. For example, nanowires of these wide band-gap materials hold promise as active materials in solar cells, light emitting diodes, and high-fidelity sensors. For most of these applications, a common desirable feature is the enhancement of the effective surface area. This can result in enhanced solar energy absorption (in solar cells) and enhanced $p-n$ junction interfaces for electron-hole pair generation (in both solar cells and LEDs). While the one-dimensional nanowire morphology naturally presents enhanced effective surface areas spanning the entire cylindrical surface, this may be further enhanced by modulating the diameter along the length of the wire. Diameter modulation demonstrated here can also lead to additional benefits such as reduced reflections from the surface thus enhancing the absorption\cite{nw:Fan:2010}, and to reduced thermal conductivity for thermoelectric applications\cite{nw:MajumdarYang:2008}.

In addition to roughness, control over the surface structure and polarity can enable novel applications. For example, Chin et al.\cite{nw:ChinAhnSunkara:2007} have observed crystal orientation dependent photoluminescence (PL) effects in GaN nanowires. Surface states are known to act as traps of photoexcited carriers. By controlling the morphology of the nanowires and their growth direction, we have tighter control of such PL phenomena that can lead to better performance of lasers or LEDs based on GaN and related compound semiconductor nanowires.

The transitions in the growth direction can lead to nanowires with engineered interfaces which can then be used as single nanowire devices. We observe transitions from the polar growth direction (serrated nanowires) to non-polar (straight nanowires) (see also Supplementary Figures S3 and S4). The transitions are associated with semi-polar segments, and the polarity-dependent anisotropic transport response  can be engineered for a range of nanoelectronic and optical devices based on individual nanowires. Further investigations of external and local growth parameters favoring this transitional growth can even lead to control over the semi-polar growth mode that remains to be realized.

In summary, this study shows a simple route based on for VLS growth of GaN nanowires with controlled growth direction, surface polarity and surface roughness. Specifically, a unique serrated morphology all through the length of the wire has been obtained for specific growth conditions. The growth morphology is attractive for a host of applications as it i) provides enhanced active surface with controlled surface polarity for specific applications, and ii) it has been achieved by controlling two  growth parameters, namely the initial catalyst size and the initial V:III ratio. A theoretical model shows that the newly discovered growth mode arises due to kinetic frustration in turn induced by energetic and geometric constraints. The model predictions are validated by the trends in the as-grown morphologies induced by systematic variations in the catalyst particle geometry and processing conditions, and the principles of shape selection are relevance for catalyzed growth of general multicomponent nanowires.

\subsection*{Methods}
The GaN nanowires were synthesized using a chemical vapor deposition (CVD) method. A $5-11$\,nm thick Au catalyst films were deposited by thermal or e-beam evaporation onto Si(111) substrates. The samples were subsequently placed in the center of a $35$\,mm ID quartz tube, in a hot-wall CVD system, about $2$\,cm downstream from the gallium source (Ga$_2$O$_3$ powder, 99.999\% purity, Alfa Aesar). The VLS growth of GaN nanowires was carried out for $1$\,hour at fixed temperature ($960\,^\circ$C) and pressure ($100$\,Torr) conditions, while flowing a mixture of NH$_3$ ($30$\,sccm, the nitrogen source) and H$_2$ ($50$ \,sccm, carrying gas) through the system. The heating and cooling were performed in an inert atmosphere with a flow rate of Ar of $70$\,sccm. The ammonia and hydrogen lines were kept open only during nanowire growth, after thermal equilibrium was reached at $960\,^\circ$C, and correspondingly the argon line was open only during heating and cooling. Separate study of the Au catalyst particle formation as a result of thin film dewetting during annealing was carried out under the same temperature and pressure conditions, but only in argon atmosphere, without the gallium and nitrogen sources, and without flowing hydrogen.

%% classes.  Rather than use \section, an appropriate macro is
%% provided that will always work.
%%%%%%%%%%%%%%%%%%%%%%%%%%%%%%%%%%%%%%%%%%%%%%%%%%%%%%%%%%%%%%%%%%%%%
\section*{Acknowledgements}
The experimental component of this work was performed under the auspices of the National Science Foundation(NSF)-ECCS Program \#0925285 (ZM, EP and LM). MU acknowledges support from NSF-DMR CMMT Program \#1106214. Certain commercial equipment, instruments, or materials are identified in this document. Such identification does not imply recommendation or endorsement by the National Institute of Standards and Technology, nor does it imply that the products identified are necessarily the best available for the purpose.

%\bibliography{references}

\pagebreak

\section*{\it Supplementary Information for\\
\bf Vapor-Liquid-Solid Growth of Serrated GaN Nanowires: Shape Selection via Kinetic Frustration\\
}

\begin{figure}[htb]
\begin{center}
\includegraphics[width=0.8\columnwidth]{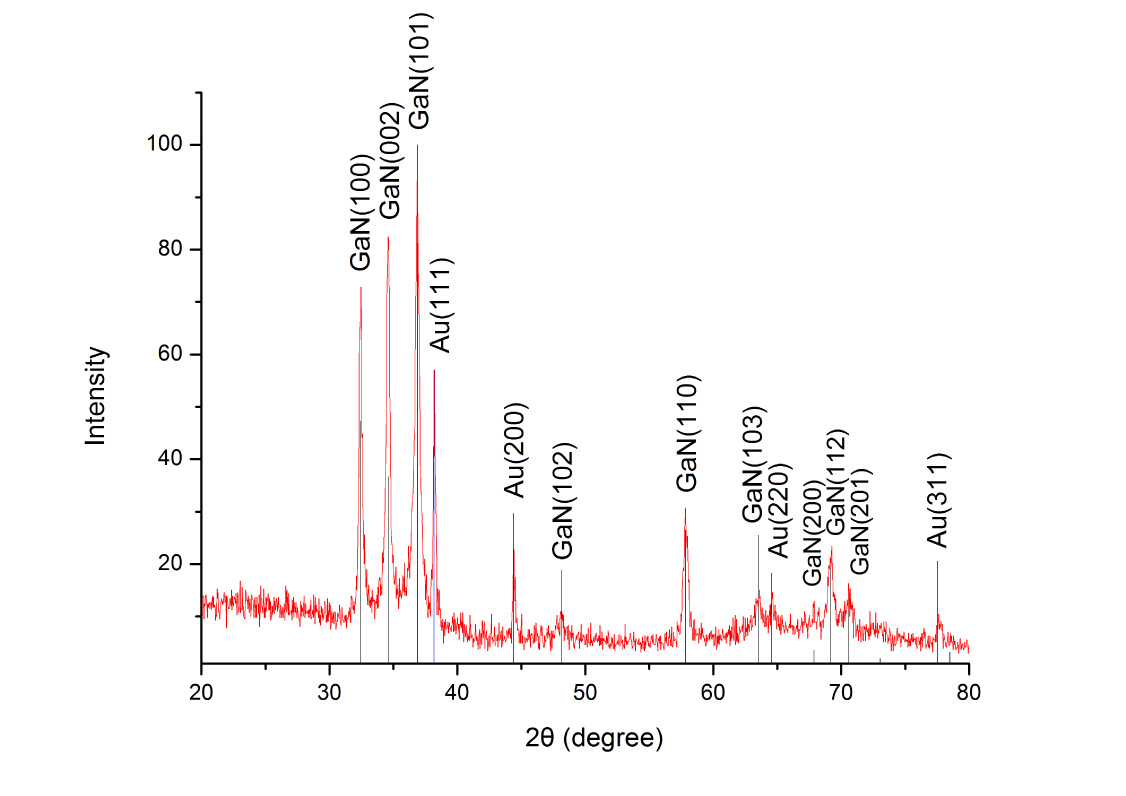}\\
\end{center}
{Figure S1: X-ray diffraction scan of an as-grown sample consisting of serrated and straight GaN nanowires indexed to the wurtzite phase with the experimental lattice parameters of $a=3.194\,{\rm \AA}$ and $c=5.196\,{\rm \AA}$. The $a$ and $c$ values indicate stress-free GaN crystal lattice~[23]. The gold peaks are due to the capping catalyst particle. 
}
\end{figure}

\pagebreak
\section*{Characterization of straight nanowires}
\begin{figure}[htb]
\begin{center}
\includegraphics[width=0.7\columnwidth]{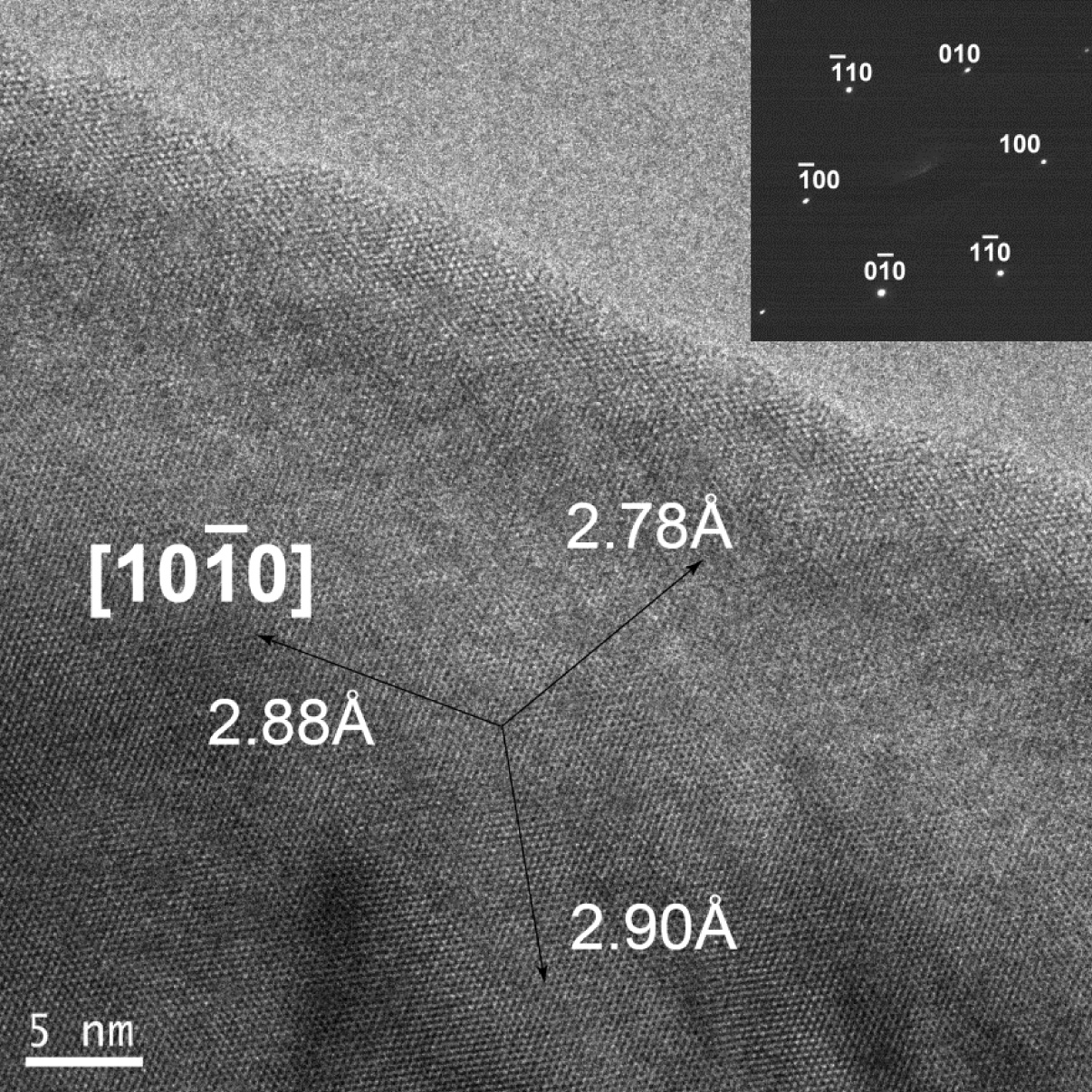}\\
\end{center}
{Figure S2: \small HRTEM image of a straight GaN nanowire viewed normal to the basal (0001) plane. The fringe spacing along the growth direction indicates non-polar growth along the $[10\bar{1}0]$ direction. 
%The interplanar spacing along the 3 equivalent directions correspond to the lattice parameter for GaN, $a=3.19$\,{\rm \AA}. 
}
\end{figure}

%\pagebreak
%\section*{Chracterization of serrated nanowires}
%\subsection*{High resolution TEM}
%\begin{figure}[h!tb]
%\centering
%\includegraphics[width=0.7\columnwidth]{FigSupp3.eps}
%\caption{HRTEM images. (a) A low resolution image showing the capping Au catalyst particle (darker contrast). (b) Detailed view of the contact line where the droplet meets the sidewall facet. (c) Detailed view of the necked region.  
%}
%\end{figure}

\pagebreak

\section*{Characterization of serrated nanowires}
\subsection*{Electron Backscatter Diffraction Analysis}
\begin{figure}[htb]
\begin{center}
\includegraphics[width=0.6\columnwidth]{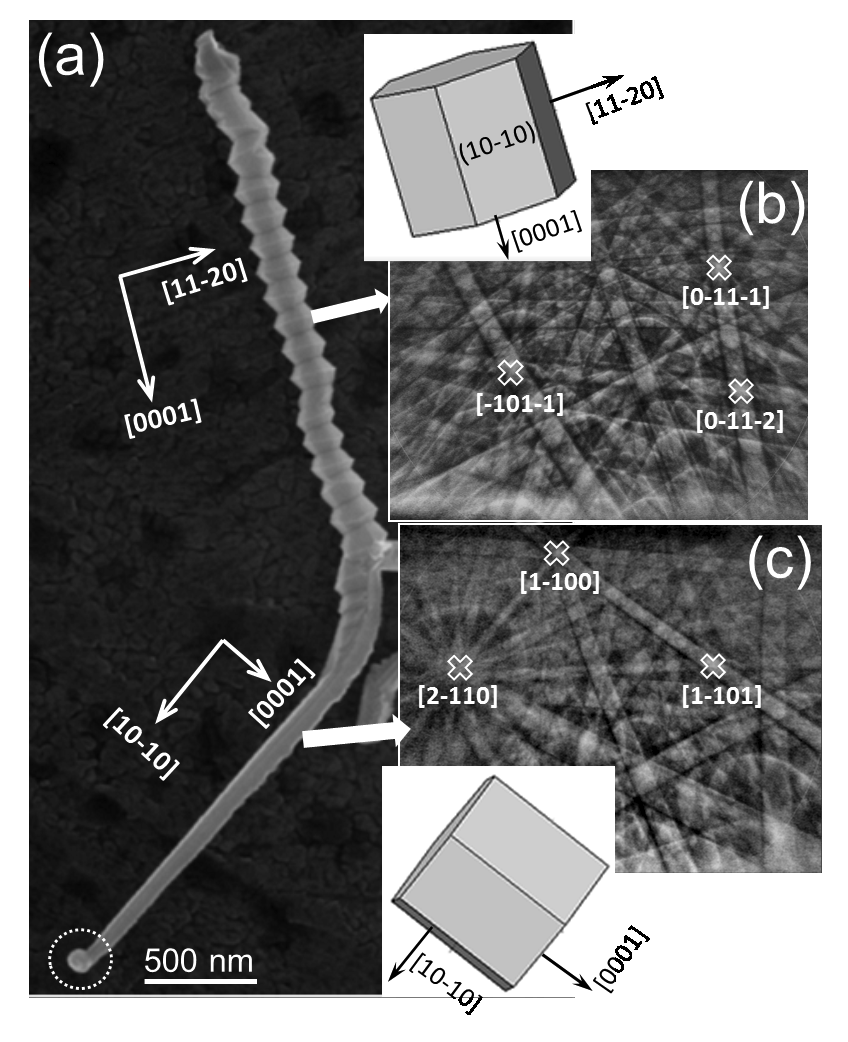}\\
\end{center}
{\small Figure S3: EBSD characterization of serrated nanowire that transitioned into straight nanowire. (a) SEM plan-view image of the whole nanowire capped by the Au catalyst particle (encircled with dashed line, bottom left). (b) EBSD pattern of the serrated segment with the simulated crystallographic orientation of the GaN unit cell (top left inset). Indexing of the EBSD pattern shows that the serrated nanowire grow along the $[0001]$-axis. (c) EBSD pattern of the straight nanowire segment with the simulated GaN unit cell (bottom left inset). The straight nanowire has a noticeably smaller size and the growth is along the non-polar $[10\bar{1}0]$ direction. 
}
\end{figure}

\pagebreak

\begin{figure}[h!tb]
\begin{center}
\includegraphics[width=0.9\columnwidth]{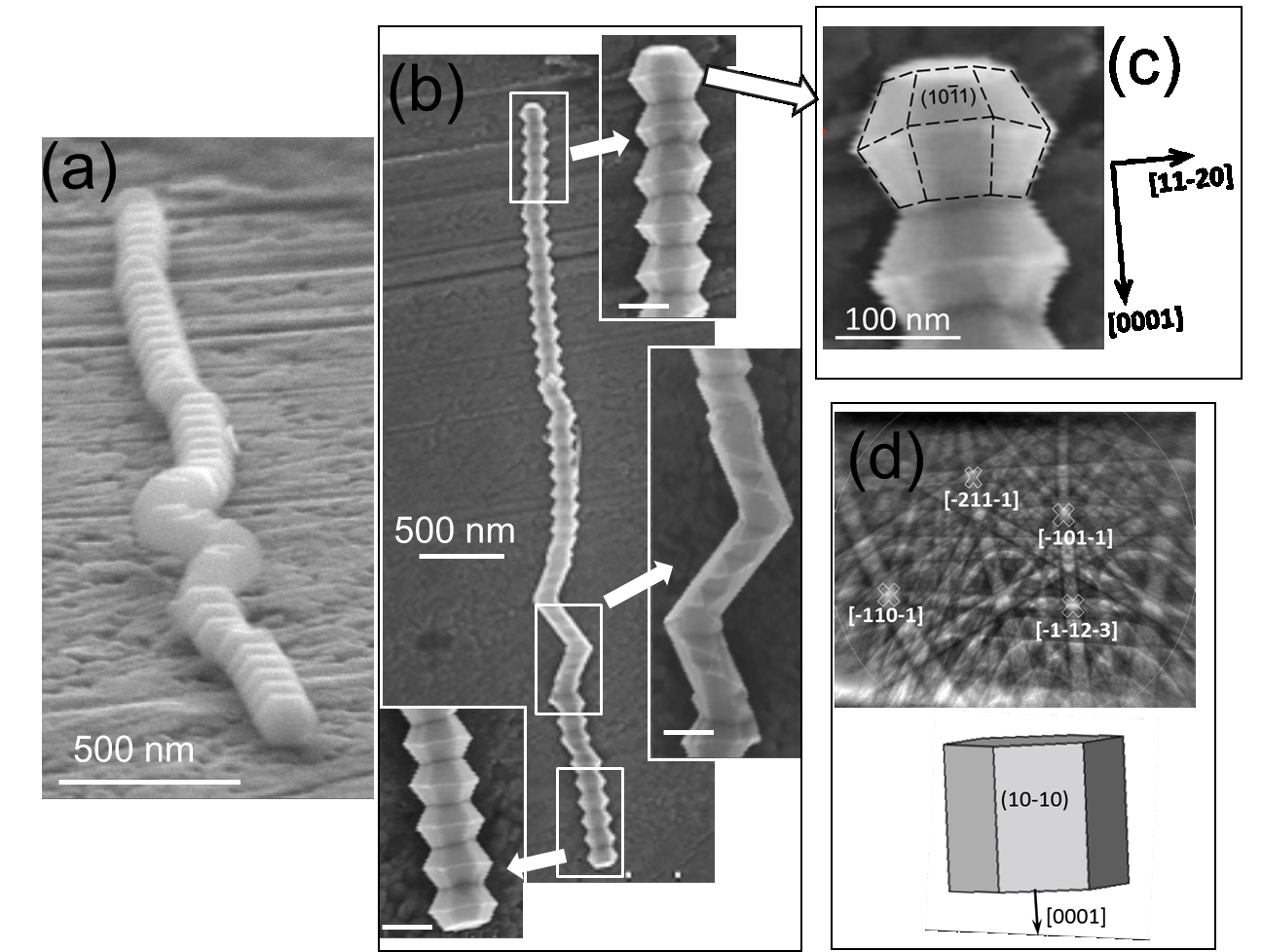}\\
\end{center}
{\small Figure S4: SEM (a)-(c) and EBSD (d) characterization of a serrated nanowire with small ``zigzaggedÓ transition zone (shown in the middle inset in (b)) that reverts back to serrated growth (lower inset in (b)). SEM $70^\circ$-tilt image of the nanowire that was harvested from the as-grown sample; 
SEM plane-view image of the same nanowire with magnified top, middle and bottom segments in the insets (scale bar in each inset is $100$\,nm)
EBSD characterization along the $[1010]$ direction identifies the growth along the polar axis enveloped by semi-polar $(1011)$ sidewall facets.
(c) High-magnification SEM of the nanowire tip with labeled $\{10\bar{1}1\}$ sidewall facets and $[0001]$ growth axis as identified from the EBSD analysis (explained below in (d)); eye-guiding dashed lines indicate six out of twelve $\{10\bar{1}1\}$ facets in the truncated hexagonal bipyramid top segment.
(d) Typical EBSD pattern from the nanowire with the simulated crystallographic orientation of the GaN unit cell. Conservation of the EBSD pattern collected from various points along the nanowire, including the serrated top, ``zigzaggedÓ transition zone, and serrated bottom segments, indicates that the nanowire preserves the $[0001]$ growth axis and $(10\bar{1}0)$ in-plane surface orientation along its whole length. The $(10\bar{1}0)$ surface orientation unambiguously defines the side facets of truncated bipyramid segments to be $\{10\bar{1}1\}$ family (as labeled in (c)).
}
\end{figure}

\pagebreak
\section*{Supplementary Notes}

\subsection*{Theoretical Model} We employ a near-equilibrium theoretical framework to relate the energetics of the nanowire-particle system to the V:III ratio at the growing nanowire front. The central aspect of Au-catalyzed GaN growth is the inherent bias in that N incorporation is limited to the contact line while Ga forms a liquid solution with Au diffuses through the particle. The relatively high growth temperature results in a liquid Au-Ga solution for sufficiently rich Ga in the gas phase. Then, the catalyst particle serves as a reservoir for Ga that changes its volume in accordance with the III:V ratio.

We begin with geometry-based scalings for the incorporation rates (atoms/time). The Ga incorporation rate scales with the exposed particle surface area, 
\begin{align}
\tag{S1}
\mathcal{I}^{Ga}=k^{Ga}_{lv} A_{lv}/\Omega^{Ga}_l \sim  k^{Ga}_{lv} R^2/\Omega^{Ga}_l,
\end{align}
where $k^{Ga}_{lv}\equiv k^{Ga}_{lv}(p_{Ga}, T)$ is the Ga catalysis rate at the droplet surface that depends on the reactor temperature and precursor gas partial pressure, $A_d$ is the droplet surface area and $\Omega^{Ga}_l$ is the Ga atomic volume within the liquid particle. The nitrogen incorporation is assumed to occur along an annular ring of width $\delta R$ around the contact line. The width is of the order of the truncating facet that is stabilized at the point where the polar [0001] facet meets the nanowire sidewalls. Then, 
\begin{align}
\tag{S2}
\mathcal{I}^{N}=k^N_c A_c/\Omega^{N}_s\sim k^N_c[R^2 - (R-\delta R)^2]/\Omega^{N}_s \approx k^N_c R\,\delta R/\Omega^{N}_s,
\end{align}
where $k^N_c \equiv k^N_c (p_{{\rm NH}_3}, T)$ is the catalysis rate associated with NH$_3$ abstraction at the contact line and $\Omega^N_s$ is the nitrogen atomic volume.

The build-up of the excess Ga and N available for nucleation and growth at the particle-nanowire interface is controlled by their respective chemical potentials. Since Ga diffuses through the liquid particle, its excess is proportional to the difference in chemical potential between the liquid particle and the solid nanowire nanowire. In the limit of small supersaturations, the chemical potentials can be simplified as
\begin{align}
\tag{S3}
\mu^{Ga}_s - \mu_{s(eq)}^{Ga} & \sim \Omega^{Ga}_s \left[\sum_i\kappa^\gamma_i + \gamma_{lv}\kappa\right]\nonumber\\ %+ \gamma_{d}\cos\theta_d)
\tag{S4}
\mu^{Ga}_l - \mu_{l(eq)}^{Ga}& \sim \ddot{G}(X^{Ga}_{l(eq)}) \left[X^{Ga}_l-X^{Ga}_{l(eq)}\right] + \Omega^{Ga}_l\gamma_{lv}\kappa . %+ 2f(\gamma^S)/R 
\end{align}
Here, the reference potentials correspond to those for the ternary equilibrium between a solid GaN and a binary (Au-Ga) alloy at the growth temperature conveniently expressed in terms of the composition of the droplet, i.e. $\mu_{s(eq)}^{Ga}=\mu_{l(eq)}^{Ga}$. The change in the solid potential is the Laplace pressure exerted by the abutting liquid particle and the contributions of the truncating and main facets. They are compactly expressed as the  weighted mean curvatures $\kappa^\gamma_i = \pm \Lambda_i/L_i$ where $\Lambda_i$ is the length (area) assigned to the facet on the Wulff plot and $L_i$ is the actual facet length (area). The sign denotes the convexity of the each facet. The expression changes at the contact line as there is an additional term due to the capillary force exerted by the liquid surface tension and the solid-vapor interface energy,  $\gamma_{d}\cos\theta_d-\gamma_{sv}\cos\alpha$, as shown in Fig.~4c. The liquid chemical potential increases linearly with the supersaturation $X^{Ga}_l-X^{Ga}_{l(eq)}$, where $X^{Ga}_{l(eq)}$ is the liquidus composition and $\ddot{G}(X^{Ga}_{l(eq)})$ is the second derivative of the free energy with respect to composition at equilibrium. 
%The term $2f(\gamma^S)/R$ is the additional Gibbs-Thomson-Herring supersaturation induced by the adjoining facet with linear dimension $R$; $f(\gamma^S$) is a function that depends on the step energy per unit height and a direct measure of the slope $d\gamma_i/d\phi$ at the cusp in the orientation-dependent solid-liquid interphase energy $\gamma_i$ [S2]. 
The last term is the classical Gibbs-Thomson term due to the Laplace pressure within the liquid. Then, the change in Ga chemical potential due to abstraction from the vapor is
\begin{align}
\tag{S5}
\Delta \mu^{Ga} & \sim \ddot{G} \left(X^{Ga}_{l(eq)}\right) \left[X^{Ga}_l-X^{Ga}_{l(eq)}\right] + \Delta\Omega^{Ga}\gamma_{d}/R + \Omega^{Ga}_s \sum_i\kappa^\gamma_i  % + 2f(\gamma^S)/R 
\end{align}
The nitrogen chemical potential of interest along contact line is set by the concentration (coverage) over the area of the surrounding annular ring that scales as $R\delta R$, 
\begin{align}
\tag{S6}
\Delta \mu^{N} & \sim \ddot{G} \left(X^{N}_{c(eq)} \right) \left(X^{N}_c -  X^{N}_{c(eq)} \right)+\gamma_{d}\sin\theta_d - \gamma_{sv}\cos\alpha\,
\end{align}
where $X^{N}_{c(eq)}$ is the equilibrium nitrogen coverage at the contact line and the last term is that due to the capillary force exerted by the liquid surface tension.

The concentration $X^{Ga}_l$ and $X^{N}$ evolve in accordance with the balance between the incoming and outgoing atomic flux (catalysis and nanowire growth, respectively). For an initially equilibrated particle-nanowire system, the steady-state values (increase per unit time) scale as
\begin{align}
\tag{S7}
\frac{d \left(X^{Ga}_l-X^{Ga}_{l(eq)}\right)}{dt} =  v\frac{d\left(\Delta X^{Ga}_l\right)}{dz}&\sim \left(\mathcal{I}^N - \frac{v R^2}{\bar{\Omega}}\right) \frac{\Omega^{Au}_l}{R^3} = \frac{\Omega^{Au}_l}{R} \left( \frac{k_{vl}^{Ga}}{\Omega^{Ga}_l} - \frac{v}{\bar{\Omega}} \right)\\
\tag{S8}
\frac{d\left(X^{N}_c -  X^{N}_{c(eq)} \right)}{dt} = v\frac{d\left(\Delta X^{N}\right)}{dz} &\sim \left( k^N_{c}- v \frac{\Omega^N_s}{\bar{\Omega}}\frac{R}{\delta R} \right),
\end{align}
where we have approximated the concentrations as the atom ratios of the relevant atomic species, i.e. $X^{Ga}_l\approx n^{Ga}/n^{Au}$ and $X^N\approx n^N/n^N_{a}$ where $n^N_a$ is the available sites along the contact line that can be occupied by nitrogen. This is reasonable due to the low solubility of Ga ($\sim5$\,at\%) and slow N catalysis under the reactor conditions.
Combining Eqs. S3-S4 and S7-S8, we arrive at scaling relations for the temporal increase in the chemical potentials within an initially equilibrated particle-nanowire upon exposure to the reactor conditions,
\begin{align}
\tag{S9}
\frac{d\left(\Delta \mu^{Ga}\right)}{dt} &\sim  \ddot{G}(X^{Ga}_{l(eq)}) \Omega^{Au}_l \left( \frac{k_{vl}}{\Omega^{Ga}_l} - \frac{v}{\bar{\Omega}} \right) - \frac{\Delta\Omega^{Ga}\gamma_{d}}{R^2}\frac{dR}{dz} + \frac{d\sum_i\kappa^\gamma_i}{dt}\\
\tag{S10}
\frac{d\left(\Delta \mu^{N}\right)}{dt} &\sim \ddot{G} \left(X^{N}_{c(eq)} \right)  \left( k^N_{ss}- v \frac{\Omega^N_s}{\bar{\Omega}}\frac{R}{\delta R} \right) + \gamma_d\cos\theta_d \frac{d\theta_d}{dt}.
\end{align}

Equations~S7-S10 capture the effect of geometry and processing conditions on the temporal increase in the concentrations and their effect on the chemical potentials of the two species. At microscopic time-scales, the growth is obviously not uniform; a critical supersaturation is required at the growth front for formation of a stable nucleus (interface-controlled regime). In the limit the catalysis is extremely slow, super-critical nucleus must wait for incorporation, and to a minor extent the diffusion, of Ga and/or N at the growth front catalysis-controlled regime which we analyze below.
%Since the nanowire growth occurs over several minutes, the growth is likely controlled by catalysis. However, for sake of completeness, both regimes are analyzed below.

\noindent
\subsection*{Catalysis-limited growth}
We consider the extreme case when the N incorporation is the rate-limiting event, i.e. the rate at which N becomes available through catalysis equals the rate at which it is absorbed into the growing nucleus, In essence, the growth velocity is kinetically limited by N-supply such that the N excess chemical potential is exactly at equilibrium and therefore constant, i.e. $d(\Delta\mu_c^N)/dt\approx0$. 
%The exposed surface becomes Ga-polar as the steady-state concentration on the exposed contact line $X^N\sim0$. 
Then, Eq.~S8 yields the steady-state growth velocity, 
\begin{align}
\tag{S11}
v=\frac{\bar{\Omega}}{\Omega^N_s}\frac{\delta R}{R}  k^N_c
%v\sim\frac{\bar{\Omega}}{\Omega^N_s}\frac{\delta R}{R}  \left( k^N_{ss} -  \frac{\gamma_{d}\cos\theta_d}{\ddot{G}(0)}  \right)  
\end{align}
Substituting in Eqs.~S7-S8, we arrive at the temporal evolution of the concentration and chemical potential of Ga,
\begin{align}
\tag{S12}
\frac{d\left(\Delta X^{Ga}\right)}{dt} &\sim \frac{\Omega^{Au}_l}{R} \left( \frac{k_{vl}^{Ga}}{\Omega^{Ga}_l} - \frac{\delta R}{R} \frac{k_{c}^N}{\Omega_s^N} \right)\\
\tag{S13}
\frac{d(\Delta \mu^{Ga})}{dt} &\sim  \frac{\ddot{G}(X^{Ga}_{l(eq)}) \Omega^{Au}_l}{R}  \left( \frac{k_{vl}^{Ga}}{\Omega^{Ga}_l} - \frac{\delta R}{R} \frac{k_{c}^N}{\Omega_s^N}  \right)
\end{align}
Ignoring the differences in the atomic volume, we recover the relations in the main text (Eqs.~1 and 2) that we reproduce for completeness,
\begin{align}
\tag{S14}
v &\sim \frac{\delta R}{R}  k^N_{c}\\
\tag{S15}
\frac{d\left(\Delta X^{Ga}\right)}{dt}  &\sim \frac{1}{R} \left(k_{vl}^{Ga} - \frac{\delta R}{R} k_{c}^N\right), \quad \frac{d\left(\Delta X^{N}\right)}{dt}\sim 0\\
\tag{S16}
\frac{d(\Delta \mu^{Ga})}{dt} &\sim  \frac{\ddot{G}(X^{Ga}_{l(eq)})}{R} \left(k_{vl} - \frac{\delta R}{R}k_{c}^N  \right) + \frac{d\sum_i\kappa^\gamma_i}{dt},\quad \frac{d\Delta \mu^{N}}{dt} \sim \gamma_d \cos\theta_d \frac{d\theta_d}{dt}.
\end{align}

\end{document}